\documentclass[12pt]{article}
  \usepackage{amsfonts}
  \usepackage{amsmath}
\usepackage{amssymb}
\usepackage{amscd}
\usepackage[dvips]{graphicx}
\begin{document}
~~
\bigskip
\bigskip
\begin{center}
{\Large {\bf{{{Twisted Rindler space-times}}}}}
\end{center}
\bigskip
\bigskip
\bigskip
\begin{center}
{{\large ${\rm {Marcin\;Daszkiewicz}}$ }}
\end{center}
\bigskip
\begin{center}
{ {{{Institute of Theoretical Physics\\ University of Wroc{\l}aw pl.
Maxa Borna 9, 50-206 Wroc{\l}aw, Poland\\ e-mail:
marcin@ift.uni.wroc.pl}}}}
\end{center}
\bigskip
\bigskip
\bigskip
\bigskip
\bigskip
\bigskip
\bigskip
\bigskip
\begin{abstract}
The (linearized) noncommutative Rindler space-times associated with
canonical, Lie-algebraic and quadratic twist-deformed Minkowski
spaces are provided. The corresponding deformed Hawking spectra
detected by Rindler observers are derived as well.
\end{abstract}
\bigskip
\bigskip
\bigskip
\bigskip
\bigskip
\bigskip
\bigskip
\bigskip
\bigskip
 \eject
\section{{{Introduction}}}

One of the most interesting discovery of modern theoretical physics
deals with  the deep relation between horizons of black hole and
thermodynamics. At the beginning of 1970s there was observed by
Bekenstein (see \cite{bekenstein}), that laws of black hole dynamics
can be provided with  thermodynamical interpretation, if one
identifies entropy with the area of black hole horizon and
temperature  with its "surface gravity". This observation has been
confirmed by Hawking in his seminal articles \cite{hawking1},
\cite{hawking2},  in which, it was predicted that a black hole
should radiate with a temperature
\begin{equation}
T_{{\rm Black\;Hole}} = \frac{\hbar g}{2\pi kc}\;, \label{tempe}
\end{equation}
where $g$ denotes the gravitational acceleration at the surface of
the black hole, $k$ is Boltzman's constant, and $c$ is the speed of
light. Subsequently, it was shown separately by Davies \cite{devies}
and Unruh \cite{unruh}, that uniformly accelerated observer in
vacuum detects a radiation (a thermal field) with the same
temperature  as $T_{{\rm Black\;Hole}}$
\begin{equation}
T_{{\rm Vacuum}} = \frac{\hbar a}{2\pi kc}\;, \label{tempe1}
\end{equation}
but with inserted   acceleration   of the detector $a$. Formally,
such an observer "lives" in so-called Rindler space-time
\cite{rindler}, which can be obtained by the following
transformation from Minkowski space with coordinates
$(x_0,x_1,x_2,x_3)$\footnote{$c=1$.}
\begin{eqnarray}
x_0 &=& N(z_1)\, {\rm sinh} (az_0)\;,\label{trans1}\\
x_1 &=& N(z_1)\, {\rm cosh} (az_0)\;,\label{trans2}\\
x_2 &=& z_2\;,\label{trans3} \\
x_3 &=& z_3\;,\label{trans4}
\end{eqnarray}
where $N$ is a positive function of the coordinate. The Minkowski
metric $ds^2 = -dx_0^2 + \sum_{i=1}^{3}dx_i^2$ transforms to
\begin{equation}
ds^2 =-aN^2(z_1)dz_0^2 + (N')^2(z_1)dz_1^2 + dz_2^2 + dz_3^2\;.
\label{metric}
\end{equation}

Recently, in \cite{krindler}, there was proposed the noncommutative
counterpart of Rindler space, so-called (linearized)
$\kappa$-Rindler
space
\begin{equation}
[\,z_0,z_{2,3}\,] = \frac{i}{az_1\kappa}\,z_{2,3}\, {\rm cosh}
(az_0)\;, \label{kapparindler}
\end{equation}
\begin{equation}
[\,z_1,z_{2,3}\,] = \frac{i}{a\kappa}\,z_{2,3}\, {\rm sinh}
(az_0)\;\;\;,\;\;\; [\,z_{2,3},z_{3,2}\,] = 0 \;,
\label{kapparindler1}
\end{equation}
associated with the well-known $\kappa$-deformed Minkowski
space-time \cite{kmin1}, \cite{kmin2}
\begin{equation}
[\,x_0,x_i\,] = \frac{i}{\kappa} x_i\;\;\;,\;\;\;\;[\,x_i,x_j\,]
=0\;\;\;;\;\;\;i,j =1,2,3\;, \label{kappaminkowski}
\end{equation}
equipped with the mass-like parameter $\kappa$. Moreover, following
the content of the papers \cite{bekenstein}-\cite{unruh} (see also
\cite{simpl1}, \cite{simpl2}), there has been found the leading
$1/\kappa$  correction to the Hawking thermal spectrum, detected by
noncommutative and uniformly accelerated ($\kappa$-Rindler)
observer.

The suggestion to use noncommutative coordinates goes back to
Heisenberg and was firstly  formalized by Snyder in \cite{snyder}.
Recently, there were also found formal  arguments based mainly  on
Quantum Gravity \cite{2}, \cite{2a} and String Theory models
\cite{recent}, \cite{string1}, indicating that space-time at Planck
scale  should be noncommutative, i.e. it should  have a quantum
nature.

Presently, in accordance with the Hopf-algebraic classification of
all deformations of relativistic and nonrelativistic symmetries
\cite{clas1}, \cite{clas2}, one can distinguish three basic types
of space-time noncommutativity:\\
\\
{\bf 1)} The canonical (soft) deformation
\begin{equation}
[\;{ x}_{\mu},{ x}_{\nu}\;] = i\theta_{\mu\nu}\;, \label{noncomm}
\end{equation}
with constant and antisymmetric tensor
$\theta_{\mu\nu}$\footnote{$\theta_{\mu\nu} = -\theta_{\nu\mu}$.}.
The explicit form of corresponding Poincare Hopf algebra has been
provided in \cite{oeckl}, \cite{chi}, while  its nonrelativistic
limit  has been
proposed in \cite{daszkiewicz}. \\
\\
{\bf 2)} The Lie-algebraic  case 
\begin{equation}
[\;{ x}_{\mu},{ x}_{\nu}\;] = i\theta_{\mu\nu}^{\rho}x_{\rho}\;,
\label{noncomm1}
\end{equation}
with  particularly chosen constant coefficients
$\theta_{\mu\nu}^{\rho}$. Particular kind of such  space-time
modification has been obtained as representations of
$\kappa$-Poincare \cite{kappaP1}, \cite{kappaP2} and
$\kappa$-Galilei \cite{kappaG} Hopf algebras. Besides, the
Lie-algebraic twist deformations of relativistic and nonrelativistic
symmetries have been provided in \cite{lie1}, \cite{lie2} and
\cite{daszkiewicz}, respectively.
\\
\\
{\bf 3)} The quadratic deformation
\begin{equation}
[\;{ x}_{\mu},{ x}_{\nu}\;] =
i\theta_{\mu\nu}^{\rho\tau}x_{\rho}x_{\tau}\;, \label{noncomm2}
\end{equation}
with constant coefficients $\theta_{\mu\nu}^{\rho\tau}$. Its
Hopf-algebraic realization was proposed in \cite{qdef}, \cite{paolo}
and \cite{lie2}  in the case of realtivistic symmetry, and in
\cite{daszkiewicz2} for its nonrelativistic counterpart.

In this article, following the scheme proposed in \cite{krindler},
we provide the noncommutative counterparts of Rindler space-time,
associated with   canonical ({\bf 1)}), Lie-algebraical ({\bf 2)}),
and quadratic ({\bf 3)}) twisted Poincare Hopf algebras,
respectively. Further, we investigate the gravito-thermodynamical
radiation detected    by such twist-deformed Rindler observers in
the vacuum, i.e.  we find the thermal (Hawking) spectra for
considered classes of twisted space-times. However, it should be
noted, that the most detailed calculation of such a spectrum has
been performed in the case of canonical deformation {\bf 1)}.

The paper is organized as follows. In  first section we recall the
basic facts concerning the twisted Poincare Hopf algebras and the
corresponding quantum space-times. The second section is devoted to
the canonically, Lie-algebraically and quadratically twisted Rindler
spaces, obtained from their noncommutative Minkowski counterparts.
The deformed Hawking radiation spectra detected by  twisted Rindler
observers are derived in  section three. The final remarks are
discussed in the last section.

\section{{{Twisted relativistic symmetries and the corresponding quantum space-times}}}

\subsection{{{Twisted Poincare Hopf algebras}}}

In this subsection we recall basic facts related with the
twist-deformed relativistic symmetries provided in \cite{chi},
\cite{lie1} and \cite{lie2}\footnote{For their nonrelativistic
counterparts see \cite{daszkiewicz}.}.

In accordance with the general  twist procedure
\cite{drin}-\cite{twist1}, the algebraic sectors for all discussed
below Hopf algebra structures remain undeformed, i.e.
$(\eta_{\mu\nu} = (-,+,+,+))$
\begin{eqnarray}
&&\left[ M_{\mu \nu },M_{\rho \sigma }\right] =i\left( \eta _{\mu
\sigma }\,M_{\nu \rho }-\eta _{\nu \sigma }\,M_{\mu \rho }+\eta
_{\nu \rho }M_{\mu
\sigma }-\eta _{\mu \rho }M_{\nu \sigma }\right) \;,  \notag \\
&&\left[ M_{\mu \nu },P_{\rho }\right] =i\left( \eta _{\nu \rho
}\,P_{\mu }-\eta _{\mu \rho }\,P_{\nu }\right) \;\;\;,\;\;\; \left[
P_{\mu },P_{\nu }\right] =0\;,  \label{nnn}
\end{eqnarray}
while the coproducts and antipodes  transform as follows
\begin{equation}
\Delta _{0}(a) \to \Delta _{\cdot }(a) = \mathcal{F}_{\cdot }\circ
\,\Delta _{0}(a)\,\circ \mathcal{F}_{\cdot
}^{-1}\;\;\;,\;\;\;
S_{\cdot}(a) =u_{\cdot }\,S_{0}(a)\,u^{-1}_{\cdot }\;,\label{fs}
\end{equation}
with $\Delta _{0}(a) = a \otimes 1 + 1 \otimes a$, $S_0(a) = -a$ and
$u_{\cdot }=\sum f_{(1)}S_0(f_{(2)})$ (we use Sweedler's notation
$\mathcal{F}_{\cdot }=\sum f_{(1)}\otimes f_{(2)}$). The  twist
element $\mathcal{F}_{\cdot } \in {\mathcal U}_{\cdot}({\mathcal P})
\otimes {\mathcal U}_{\cdot}({\mathcal P})$, present in the above
formula,  satisfies the classical cocycle  condition 
\begin{equation}
{\mathcal F}_{{\cdot }12} \cdot(\Delta_{0} \otimes 1) ~{\cal
F}_{\cdot } = {\mathcal F}_{{\cdot }23} \cdot(1\otimes \Delta_{0})
~{\mathcal F}_{{\cdot }}\;, \label{cocyclef}
\end{equation}
and the normalization condition
\begin{equation}
(\epsilon \otimes 1)~{\cal F}_{{\cdot }} = (1 \otimes
\epsilon)~{\cal F}_{{\cdot }} = 1\;, \label{normalizationhh}
\end{equation}
with ${\cal F}_{{\cdot }12} = {\cal F}_{{\cdot }}\otimes 1$ and
${\cal F}_{{\cdot }23} = 1 \otimes {\cal F}_{{\cdot }}$\footnote{In
this article we consider only  Abelian twist deformations of
Poincare Hopf structures (see \cite{drin}-\cite{twist1}).}.

Let us start with the first, canonically twisted Poincare Hopf
algebra ${\mathcal U}_{\theta_{\mu\nu}}({\mathcal P})$ provided in
\cite{chi}. Its algebraic part remains classical (see formula
(\ref{nnn})), while the corresponding twist factor and coproducts
take the forms
\begin{equation}
 {\mathcal{F}}_{\theta }=\exp
i\left(\theta^{\mu\nu}P_{\mu }\wedge P_{\nu} \right)\;\;\;;\;\;\; a
\wedge b = a \otimes b -b \otimes a\;, \label{gfactor}
\end{equation}
and
\begin{eqnarray}
\Delta_{\theta_{\mu\nu}}(P_\rho)&=&\Delta_0(P_\rho)\;, \label{dlww3v}\\
\Delta _{\theta_{kl} }(M_{\mu \nu }) &=&\Delta _{0}(M_{\mu \nu })-
\theta_{kl}[(\eta _{k \mu }P_{\nu }-\eta _{k \nu }\,P_{\mu })\otimes
P_{l }+P_{k}\otimes (\eta_{l
\mu}P_{\nu}-\eta_{l \nu}P_{\mu})]\nonumber\\
&+& \theta_{kl}[(\eta _{l \mu }P_{\nu }-\eta _{l \nu }\,P_{\mu
})\otimes P_{k }+P_{l}\otimes (\eta_{k \mu}P_{\nu}-\eta_{k
\nu}P_{\mu})]\;.\label{zadruzny}
\end{eqnarray}
 Of course, for deformation parameter  $\theta_{\mu\nu}$
approaching zero we get the undeformed (classical) Poincare Hopf
structure ${\mathcal U}_{0}({\mathcal P})$.

The second twist-deformed Poincare Hopf algebra
$\mathcal{U}_{\kappa}(\mathcal{P})$ has been provided in \cite{lie1}
and \cite{lie2}. Its algebraic part remains classical while
coproducts take the form
\begin{eqnarray}
 \Delta_\kappa(P_\mu)&=&\Delta
_0(P_\mu)+(-i)^\gamma\sinh \left(
\frac{i^{\gamma}}{2\kappa}\zeta^\lambda P_\lambda \right)\wedge
\left(\eta_{\alpha \mu}P_\beta -\eta_{\beta \mu}P_\alpha \right)\label{swiatowid100}\\
&+&(\cosh \left(\frac{i^{\gamma}}{2\kappa}\zeta^\lambda  P_\lambda
\right)-1)\perp \left(\eta_{\alpha \alpha }\eta_{\alpha \mu}P_\alpha
+\eta_{\beta \beta }\eta_{\beta \mu}P_\beta \right)\;,
\notag  \\
&~~&  \cr \Delta_\kappa(M_{\mu\nu})&=&\Delta_0(M_{\mu\nu})+M_{\alpha
\beta }\wedge \frac{1}{2\kappa}\zeta^\lambda \left(\eta_{\mu
\lambda }P_\nu-\eta_{\nu \lambda }P_\mu\right)\label{swiatowid200}\\
&+&i\left[M_{\mu\nu},M_{\alpha \beta }\right]\wedge
(-i)^{\gamma}\sinh\left(\frac{i^{\gamma}}{2\kappa}\zeta^\lambda P_\lambda \right) \notag \\
&+&\left[\left[%
M_{\mu\nu},M_{\alpha \beta }\right],M_{\alpha \beta
}\right]\perp(-1)^{1+\gamma}
(\cosh\left(\frac{i^{\gamma}}{2\kappa}\zeta^\lambda  P_\lambda  \right)-1)  \nonumber \\
&+&M_{\alpha \beta
}(-i)^\gamma\sinh\left(\frac{i^{\gamma}}{2\kappa}\zeta^\lambda
P_\lambda \right)\perp
\frac{1}{2\kappa}\zeta^\lambda \left(\psi_\lambda P_\alpha -\chi_\lambda P_\beta \right) \notag \\
&+&\frac{1}{2\kappa}\zeta^\lambda \left(\psi_\lambda \eta_{\alpha
\alpha }P_\beta +\chi_\lambda \eta_{\beta \beta }P_\alpha
\right)\wedge M_{\alpha \beta
}(-1)^{1+\gamma}(\cosh\left(\frac{i^{\gamma}}{2\kappa}\zeta^\lambda
P_\lambda \right)-1)\;,
  \notag
\end{eqnarray}
 where  $a\perp b=a\otimes
b+b\otimes a$, $\psi_\gamma =\eta_{j \gamma }\eta_{l i}-\eta_{i
\gamma }\eta_{lj},\; \chi_\gamma =\eta_{j \gamma }\eta_{k i}-\eta_{i
\gamma }\eta_{k j}$ and $\zeta^\lambda$ - arbitrary fourvector. The
corresponding twist factor looks as follows
\begin{equation}
 {\mathcal{F}}_{\kappa }= \exp \frac{i}{2\kappa}(
\zeta^\lambda\,P_{\lambda }\wedge M_{\alpha \beta
})\;\;\;;\;\;\;\lambda \neq \alpha,\beta \;. \label{ggfactor}
\end{equation}
Obviously, for deformation parameter $\kappa$ running to infinity
the above Hopf structure becomes classical.

The last twist deformation of relativistic symmetries, as we shall
see below, generates the quadratic space-time noncommutativity, and
is associated with the following twist factor
\begin{equation}
 {\mathcal{F}}_{\xi }= \exp\frac{i}{2}\xi\,(
M_{\alpha\beta}\wedge
M_{\gamma\delta})\;\;\;;\;\;\;\alpha,\beta,\gamma,\delta\,-\,{\rm
different\;and\;fixed} \;. \label{gggfactor}
\end{equation}
This type of deformation has been proposed in \cite{lie2} but,
unfortunately, due to the nontrivial technical problems, the
explicit form of its coalgebraic sector  has not been provided in
explicite form.

\subsection{{{Twisted Minkowski space-times}}}

In this subsection we introduce the generalized relativistic
space-times corresponding to the Poincare Hopf algebras provided in
the  pervious section. They are defined as  quantum representation
spaces (Hopf modules) for quantum Poincare algebras, with action of
the deformed symmetry generators satisfying suitably deformed
Leibnitz rules \cite{bloch}, \cite{3b}, \cite{chi}. The action of
Poincare algebra on its Hopf module of functions depending on
space-time coordinates ${x}_\mu$ is given by

\begin{equation}
P_{\mu }\rhd f(x)=i\partial _{\mu }f(x)\;\;\;,\;\;\; M_{\mu \nu
}\rhd f(x)=i\left( x_{\mu }\partial _{\nu }-x_{\nu }\partial _{\mu
}\right) f(x)\;,  \label{a1}
\end{equation}
while the $\star_{.}$-multiplication of arbitrary two functions  is
defined as follows
\begin{equation}
f({x})\star_{{\cdot}} g({x}):= \omega\circ\left(
 \mathcal{F}_{\cdot}^{-1}\rhd  f({x})\otimes g({x})\right) \;.
\label{star}
\end{equation}
In the above formula $\mathcal{F}_{\cdot}$ denotes  twist factor
corresponding to the proper Poincare algebra  and $\omega\circ\left(
a\otimes b\right) = a\cdot b$.

Hence, we get the following twisted Minkowski
space-times:\footnote{$[\;a,b\;]_{\star} :=
a\star b - b\star a\;. $}\\
\\
$i)$ Canonical deformation of relativistic space
\begin{equation}
[\;{ x}_{\mu},{ x}_{\nu}\;]_{\star_{\theta} } =
i\theta_{\mu\nu}\;,
\label{wielkaslawia}
\end{equation}
$ii)$ Lie-algebraically deformed Minkowski space-time
\begin{equation}
\left[ x_{\mu },x_{\nu }\right]_{\star_{\kappa} }=iC_{\ \mu \nu
}^{\rho }x_{\rho }\;, \label{aalie}
\end{equation}
with coefficients
\begin{equation}
C_{\ \mu \nu }^{\rho} = \frac{1}{\kappa}\zeta_{\mu}
\left(\eta_{\beta \nu}\delta^{\rho}_{\,\alpha}
-\eta_{\alpha\nu}\delta^{\rho}_{\,\beta} \right) +
\frac{1}{\kappa}\zeta_{\nu} \left(\eta_{\alpha
\mu}\delta^{\rho}_{\,\beta} -\eta_{\beta\mu}\delta^{\rho}_{\,\alpha}
\right) \;, \label{coef}
\end{equation}
and\\
\\
 $iii)$ Quadratic modification of relativistic space
\begin{equation}
\left[ x_{\mu },x_{\nu }\right] _{\star_\xi }=iC_{\ \mu \nu }^{\rho
\tau }x_{\rho }x_{\tau }\;, \label{aalie}
\end{equation}
where $C_{\ \mu \nu }^{\rho \tau } = C_{\ \mu \nu }^{\rho \tau
}(\xi) $ denotes the proper function of deformation parameter $\xi$,
such that\footnote{$\{\,a,b\,\}_{\star_\xi}:= a{\star_\xi} b+
b{\star_\xi}a.$}
\begin{eqnarray}
\left[ x_{\mu },x_{\nu }\right] _{\star_\xi }&=&i {\rm
tanh}\left(\frac{\xi}{2}\right)
\left(\eta_{\alpha\mu}\eta_{\gamma\nu}\{\,x_\beta,x_\delta\,\}_{\star_\xi}
-\eta_{\alpha\mu}\eta_{\delta\nu}\{\,x_\beta,x_\gamma\,\}_{\star_\xi}
+\right.\\ \nonumber &-&
\left.\eta_{\beta\mu}\eta_{\gamma\nu}\{\,x_\alpha,x_\delta\,\}_{\star_\xi}
+\eta_{\beta\mu}\eta_{\delta\nu}\{\,x_\alpha,x_\gamma\,\}_{\star_\xi}
\right) \;. \label{5009}
\end{eqnarray}
\\
 Of course, if parameters $\theta^{\mu\nu}$ and $\xi$
goes  to zero and parameter $\kappa$ approaches  infinity, the above
space-times become classical Minkowski space.

\section{{{Twisted Rindler space-times}}}

Let us now introduce  the  new objects - the twisted Rindler spaces
-  corresponding to the twisted Minkowski space-times described in
pervious section. Following  the scheme proposed in the case of
$\kappa$-Poincare deformation \cite{krindler}, we define such
space-times as the quantum spaces  with noncommutativity given by
the proper $\ast$-multiplications. Similarly to the twisted
Minkowski space-time the  new $\ast$-multiplications are defined by
the new $\mathcal{Z}$-factors, which play the role of  counterparts
of  factors (\ref{gfactor}), (\ref{ggfactor}) and (\ref{gggfactor}).
Hence, first of all,
 we recall the standard
transformation rules from commutative Minkowski space (described by
$x_\mu$ variables) to the accelerated and commutative as well
(Rindler) space-time ($z_\mu$) \cite{rindler}
\begin{eqnarray}
x_0 &=& z_1\, {\rm sinh} (az_0)\;,\label{strans1}\\
x_1 &=& z_1\, {\rm cosh} (az_0)\;,\label{strans2}\\
x_2 &=& z_2\;,\label{strans3} \\
x_3 &=& z_3\;,\label{strans4}
\end{eqnarray}
where $a$ denotes the acceleration parameter, i.e. we have chosen
function $N(z_1) = z_1$ in formulas (\ref{trans1})-(\ref{trans4}).
Next, we  rewrite  the Minkowski twist factors (\ref{gfactor}),
(\ref{ggfactor}) and (\ref{gggfactor}) (depending on commutative
$x_\mu$ variables and defining the $\star$-multiplication
(\ref{star})) in terms of
$z_\mu$ variables. In such a way, we get the Rindler $\mathcal{Z}$-factors and,   consequently, we have:\\
\\
$i)$ Canonically deformed Rindler space-time.

In such a case, due to the transformation  rules
(\ref{strans1})-(\ref{strans4})\footnote{By straightforward but
tedious calculations, we find ${\partial_{ x_0}} = (-{\rm
sinh}(az_0){\partial_{ z_1}} + ({{\rm
cosh}(az_0)}/{az_1}){\partial_{ z_0}})$ and ${\partial_{ x_1}} =
({\rm cosh}(az_0){\partial_{ z_1}} - ({{\rm
sinh}(az_0)}{az_1}){\partial_{ z_0}})$.}, the wanted
$\ast_{\theta}$-product
 takes the form
\begin{equation}
f({z})\ast_{{\theta}} g({z}):= \omega\circ\left(
 \mathcal{Z}_{\theta}^{-1}\rhd  f({z})\otimes g({z})\right) \;,
\label{rstar}
\end{equation}
where
\begin{eqnarray}
{\mathcal{Z}}_{\theta }^{-1} &=& \exp
-2i\left(\theta^{01}f_{0}(z_0,z_1,\partial_{z_0},\partial_{z_1})
\wedge f_{1}(z_0,z_1,\partial_{z_0},\partial_{z_1})+
\sum_{a=2}^3\theta^{0a} \right.\nonumber \\
&+&\left.f_{0}(z_0,z_1,\partial_{z_0},\partial_{z_1})\wedge
\partial_{z_a} - \theta^{23}\partial_{z_2} \wedge \partial_{z_3} + \right.
\label{wielkaslawia200} \\
&-&\left.\sum_{a=2}^3\theta^{1a}f_{1}(z_0,z_1,\partial_{z_0},\partial_{z_1})\wedge
\partial_{z_a}
\right) = \exp {\cal O}_{\theta}(z,\partial_z)\;, \nonumber
\end{eqnarray}
and
\begin{eqnarray}
f_{0}(z_0,z_1,\partial_{z_0},\partial_{z_1}) &=& -{\rm
sinh}(az_0)i{\partial_{ z_1}} + ({{\rm
cosh}(az_0)}/{az_1})i{\partial_{ z_0}}\;,\label{funkcja0}\\
f_{1}(z_0,z_1,\partial_{z_0},\partial_{z_1}) &=& {\rm
cosh}(az_0)i{\partial_{ z_1}} - ({{\rm
sinh}(az_0)}/{az_1})i{\partial_{ z_0}}\;.\label{funkcja1}
\end{eqnarray}
However, to simplify, we consider the following differential
operator
\begin{equation}
({ {\mathcal{Z}}}_{\theta }^{\rm Linear})^{-1} =  1+{\cal
O}_{\theta}(z,\partial_z)\;,\label{linear}
\end{equation}
which  contains only the  terms linear in deformation parameter
$\theta^{\mu\nu}$\footnote{As we shall see in the next section, we
 look for the corrections   to
Hawking radiation linear in deformation parameter.}. Hence, the
linearized ${\hat \ast}$-Rindler multiplication is given by the
formula (\ref{rstar}), but with differential operator (\ref{linear})
instead the complete  factor ${\mathcal{Z}}_{\theta }^{-1}$.
Consequently, for $f(z) = z_\mu$ and $g(z) = z_\nu$, we get
\begin{equation}
[\;{ z}_{\mu},{ z}_{\nu}\;]_{{\hat \ast}_{\theta}} =
-2i\theta^{\rho\tau}\left[(f_\rho(z,\partial_z)z_\mu)(f_\tau(z,\partial_z)z_\nu)
-(f_\tau(z,\partial_z)z_\mu)(f_\rho(z,\partial_z)z_\nu)
\right]\;,
\label{canonrindler}
\end{equation}
with $f_2(z,\partial_z) = i\partial_{z_2}$, $f_3(z,\partial_z)=
i\partial_{z_3}$. The above commutation relations define the
canonically twisted Rindler space-time associated with canonical
Minkowski space
(\ref{wielkaslawia}). \\
\\
$ii)$ Lie-algebraically deformed Rindler space.

Here, due to the rules (\ref{strans1})-(\ref{strans4}), the
$\ast_{\kappa}$-multiplication look as  follows
\begin{equation}
f({z})\ast_{\kappa} g({z})= \omega\circ\left(
 \mathcal{Z}_{\kappa}^{-1}\rhd  f({z})\otimes g({z})\right) \;,
\label{rstar2}
\end{equation}
where  $(\delta_{ab/c} = \delta_{ab}$ or $\delta_{ac})$
\begin{eqnarray}
{\mathcal{Z}}_{\kappa }^{-1} &=& \exp
\frac{-i}{2\kappa}\zeta^\lambda\left((\delta_{\lambda
2/3}i\partial_\lambda + \delta_{\lambda
0/1}f_{\lambda}(z_0,z_1,\partial_{z_0},\partial_{z_1}))\;
\wedge\right.\nonumber\\
&\wedge& \left. ((\delta_{\alpha2/3}z_\alpha + \delta_{\alpha
0/1}g_\alpha(z_0,z_1))(\delta_{\beta2/3}i\partial_\beta +
\delta_{\beta 0/1}f_{\beta}(z_0,z_1,\partial_{z_0},\partial_{z_1})
)\right.\label{wielkaslawia201}\\
&-&\left.(\delta_{\beta2/3}z_\beta + \delta_{\beta
0/1}g_\beta(z_0,z_1))(\delta_{\alpha2/3}i\partial_\alpha +
\delta_{\alpha
0/1}f_{\alpha}(z_0,z_1,\partial_{z_0},\partial_{z_1}))\right) \nonumber\\
&=& = \exp \frac{-i}{2\kappa}\zeta^\lambda\left( {\cal
A}_{\lambda}(z,\partial_z)\wedge {\cal
A}_{\alpha\beta}(z,\partial_z) \right) = \exp {\cal
O}_{\kappa}(z,\partial_z)\;,\nonumber
\end{eqnarray}
 and
\begin{equation}
g_{0}(z_0,z_1) =z_1\, {\rm sinh} (az_0) \;\;\;,\;\;\; g_{1}(z_0,z_1)
=z_1\, {\rm cosh} (az_0) \;.\label{gfunkcje}
\end{equation}
Consequently, the corresponding (linearized) Rindler space-time
takes the form
\begin{equation}
[\;{ z}_{\mu},{ z}_{\nu}\;]_{{\hat \ast}_{\kappa}}
=-\frac{i}{\kappa}\zeta^\lambda\left[({\cal
A}_{\lambda}(z,\partial_z)z_\mu)({\cal
A}_{\alpha\beta}(z,\partial_z)z_\nu) -({\cal
A}_{\alpha\beta}(z,\partial_z)z_\mu)({\cal
A}_{\lambda}(z,\partial_z)z_\nu) \right]
\;,
\label{lierindler}
\end{equation}
where we use  the linearized approximation to
${{\mathcal{Z}}}_{\kappa }^{-1}$ (see (\ref{linear})).\\
\\
$iii)$ Quadratic  deformation of Rindler space.

In such a  case, the $\ast_{\xi}$-multiplication  takes the form
\begin{equation}
f({z})\ast_{\xi} g({z})= \omega\circ\left(
 \mathcal{Z}_{\xi}^{-1}\rhd  f({z})\otimes g({z})\right) \;,
\label{rstar3}
\end{equation}
with factor
\begin{eqnarray}
{\mathcal{Z}}_{\xi }^{-1}&=&
\exp\frac{-i}{2}\xi\left(((\delta_{\alpha2/3}z_\alpha +
\delta_{\alpha
0/1}g_\alpha(z_0,z_1))(\delta_{\beta2/3}i\partial_\beta +
\delta_{\beta 0/1}f_{\beta}(z_0,z_1,\partial_{z_0},\partial_{z_1})
\right.\nonumber\\
&-&\left.(\delta_{\beta2/3}z_\beta + \delta_{\beta
0/1}g_\beta(z_0,z_1))(\delta_{\alpha2/3}i\partial_\alpha +
\delta_{\alpha
0/1}f_{\alpha}(z_0,z_1,\partial_{z_0},\partial_{z_1})))\;
\wedge\right.\nonumber\\
&\wedge& \left.( (\delta_{\gamma2/3}z_\gamma + \delta_{\gamma
0/1}g_\gamma(z_0,z_1))(\delta_{\delta2/3}i\partial_\delta +
\delta_{\delta
0/1}f_{\delta}(z_0,z_1,\partial_{z_0},\partial_{z_1}))\;+
\right.\label{twistkwad}\\
&-&\left.(\delta_{\delta2/3}z_\delta + \delta_{\delta
0/1}g_\delta(z_0,z_1))(\delta_{\gamma2/3}i\partial_\gamma +
\delta_{\gamma
0/1}f_{\gamma}(z_0,z_1,\partial_{z_0},\partial_{z_1})))\right)\nonumber\\
&=& \exp\frac{-i}{2}\xi\left({\cal
A}_{\alpha\beta}(z,\partial_z))\wedge {\cal
A}_{\gamma\delta}(z,\partial_z))\right) = \exp {\cal
O}_{\xi}(z,\partial_z)\;.\nonumber
\end{eqnarray}
Then,  the (linearized) quadratically deformed Rindler space looks
as follows
\begin{equation}
[\;{ z}_{\mu},{ z}_{\nu}\;]_{{\hat \ast}_{\xi}} = -i\xi \left[({\cal
A}_{\alpha\beta}(z,\partial_z)z_\mu)({\cal
A}_{\gamma\delta}(z,\partial_z)z_\nu) -({\cal
A}_{\gamma\delta}(z,\partial_z)z_\mu)({\cal
A}_{\alpha\beta}(z,\partial_z)z_\nu) \right]
\;,
\label{quarindler}
\end{equation}
with ${{\hat \ast}_{\xi}}$-multiplication defined by the linear
approximation to (\ref{twistkwad}) (see (\ref{linear})).\\
\\
Obviously, for both deformation parameters $\theta^{\mu\nu}$ and
$\xi$ approaching zero, and parameter $\kappa$ running to infinity,
the above (twisted) Rindler space-times become classical.

\section{{{Hawking thermal spectra  for  twisted Rindler space-times}}}

In this   section we find the corrections to the
gravito-thermodynamical process, which occur  in twisted
(noncommutative) space-times. It should be noted, however, that the
more detailed calculations of a proper spectrum have been  performed
only in the case of canonical deformation {{\bf 1)}}.

As it was mentioned in Introduction, such effects as Hawking
radiation \cite{hawking1},  can be observed in vacuum by uniformly
accelerated observer \cite{devies}, \cite{unruh}. First of all,
following \cite{krindler}, we  recall the calculations performed for
 gravito-thermodynamical process in commutative relativistic space-time
\cite{simpl1}, \cite{simpl2}. Firstly, we consider the on-shell
plane wave corresponding to the massless mode with positive
frequency $\hat{\omega}$ moving in $x_1=x$ direction of Minkowski
space ($x_0 =t$)
\begin{equation}
\phi (x,t) = \exp \left(\hat{\omega}x- \hat{\omega}t \right) \;.
\label{field1}
\end{equation}
In terms of Rindler variables this plane wave takes the form
$(z_0=\tau, z_1=z)$
\begin{equation}
\phi (x(z,\tau),t(z,\tau)) \equiv \phi (z,\tau)  = \exp
\left(i\hat{\omega}z {\rm e}^{-a\tau} \right) \;, \label{field2}
\end{equation}
i.e. it  becomes nonmonochromatic  and instead has the frequency
spectrum $f(\omega)$, given by Fourier transform
\begin{equation}
 \phi (z,\tau) = \int_{-\infty}^{+\infty}\frac{d\omega}{2\pi}
 f(\omega){\rm
 e}^{-i\omega \tau}
\;.  \label{trans}
\end{equation}
The corresponding power spectrum is given by $P(\omega) =
|f(\omega)|^2$ and the function $f(\omega)$ can be obtained  by
inverse Fourier transform
\begin{equation}
f(\omega)  =\int_{-\infty}^{+\infty}d\tau\,{\rm e}^{i\hat{\omega}z
{\rm e}^{-a\tau}}{\rm
 e}^{i\omega \tau} = \left(-\frac{1}{a}\right)\left({\hat \omega}z\right)^{i\omega/a}\Gamma
 \left(-\frac{i\omega}{a}\right){\rm e}^{\pi\omega/2a}
\;,  \label{invtrans}
\end{equation}
where $\Gamma (x)$ denotes the gamma function \cite{gamma}. Then,
since
\begin{equation}
\left|\Gamma\left(\frac{i{ \omega}}{a}\right)\right|^2  =
\frac{\pi}{({ \omega}/a){\rm sinh}(\pi{ \omega}/a)} \;,
\label{gamma}
\end{equation}
we get the following power spectrum at negative frequency
\begin{equation}
\omega P(-\omega) = \omega |f(-\omega)|^2  = \frac{2\pi/a}{{\rm
e}^{2\pi\omega/a}-1}\;, \label{power}
\end{equation}
which corresponds to the  Planck factor $\left({\rm
e}^{{\hbar\omega}/kT} -1 \right)$ associated with temperature $T =
\hbar a/2\pi kc$ (the temperature of radiation seen by Rindler
observer (see formula (\ref{tempe1}))).

Let us now turn to the case of twisted (noncommutative) space-times
provided in pervious section. In order to find the power spectra for
such deformed Rindler spaces, we start with the (fundamental)
formula (\ref{field2}) for scalar field, equipped with the twisted
(linearized) ${\hat \ast}$-multiplications
\begin{equation}
 \phi^{{\rm Twisted}}_{\cdot} (z,\tau)  = \exp
\left(i\hat{\omega}z{\hat \ast}_{\cdot} {\rm e}^{-a\tau} \right) \;.
\label{field222}
\end{equation}
Then
\begin{equation}
f^{{\rm Twisted}}_{\cdot}(\omega)  =
\int_{-\infty}^{+\infty}d\tau\,{\rm e}^{i\hat{\omega}z{\hat
\ast}_{\cdot} {\rm e}^{-a\tau}} {\hat \ast}_{\cdot} {\rm
 e}^{i\omega \tau}
\;, \label{modyfication}
\end{equation}
and, in accordance with the pervious  considerations, we get\footnote{We only take under consideration the terms linear in deformation parameters.}
\begin{eqnarray}
f^{{\rm Twisted}}_{\cdot}(\omega)  &=& f(\omega) +
\int_{-\infty}^{+\infty}d\tau\, \omega\circ\left(
 {\cal
O}_{\cdot}(\tau,z,\partial_\tau,\partial_z)\rhd  {\rm
e}^{i\hat{\omega}z {\rm e}^{-a\tau}}\otimes {\rm
 e}^{i\omega \tau}\right) +   \label{modyfication1}\\
 &+&
\int_{-\infty}^{+\infty}d\tau\, {\rm e}^{i\hat{\omega}z {\rm
e}^{-a\tau}} {\rm
 e}^{i\omega \tau}\omega\circ\left(
 {\cal
O}_{\cdot}(\tau,z,\partial_\tau,\partial_z)\rhd i{\hat \omega} z
\otimes {\rm e}^{-a\tau} \right) = f(\omega) + f_{\cdot}(\omega)\;.
\nonumber
\end{eqnarray}
Consequently, the power spectrum at negative frequency takes the
form (see e.g. \cite{simpl1}, \cite{simpl2})
\begin{eqnarray}
\omega P^{{\rm Twisted}}_{\cdot}(-\omega) &=& \omega\left|f^{{\rm
Twisted}}_{\cdot}(-\omega) \right|^2 = \omega P(-\omega) + \omega
P_{\cdot}(-\omega) =  \label{modyfication2}\\
&=& \frac{2\pi/a}{{\rm e}^{2\pi\omega/a}-1}+ \omega
P_{\cdot}(-\omega)\;,\nonumber
\end{eqnarray}
with the corrections given by\footnote{${\bar z}$ denotes complex
conjugation  for complex number $z$.}
\begin{equation}
P_{\cdot}(-\omega) = \left|f_{\cdot}(-\omega) \right|^2 + {\bar
f}_{\cdot}(-\omega)f(-\omega) + {f}_{\cdot}(-\omega){\bar
f}(-\omega) \;. \label{modyficationppp}
\end{equation}

As an example, let us consider the most simple  case of space-time
noncommutativity - the canonical deformation of classical space {\bf
1)}. In such a case the operator ${\cal
O}_{\cdot}(\tau,z,\partial_\tau,\partial_z) = {\cal
O}_{\theta}(\tau,z,\partial_\tau,\partial_z)$ is given by the
formula (\ref{wielkaslawia200}) and,  we get
\begin{equation}
f_{\theta}(\omega) = \frac{2i\theta^{01}\omega{\hat
\omega}}{az}\int_{-\infty}^{+\infty}d\tau\,{\rm e}^{i\hat{\omega}z
{\rm e}^{-a\tau}} {\rm
 e}^{i\omega \tau}{\rm e}^{-a\tau}
-\frac{2\theta^{01}{\hat
\omega}}{z}\int_{-\infty}^{+\infty}d\tau\,{\rm e}^{i\hat{\omega}z
{\rm e}^{-a\tau}} {\rm
 e}^{i\omega \tau}{\rm e}^{-a\tau}
 \;.
\label{modyficationp}
\end{equation}
Consequently, in accordance with (\ref{modyfication1}) one obtains
\begin{equation}
f^{{\rm Twisted}}_{\theta}(\omega)  =
\int_{-\infty}^{+\infty}d\tau\,{\rm e}^{i\hat{\omega}z {\rm
e}^{-a\tau}} {\rm
 e}^{i\omega \tau}\left(1+ \frac{2\theta^{01}{\hat
\omega}}{z}{\rm e}^{-a\tau}\left[\frac{i\omega}{a} -1 \right]
\right)\;.\label{modyp}
\end{equation}
The above integral can be   evaluated with the help of standard
identity for Gamma function $ \Gamma(y+1) = y\Gamma (y)$; one gets
\begin{equation}
f^{{\rm Twisted}}_{\theta}(\omega)  =
\left(-\frac{1}{a}\right)\left({\hat
\omega}z\right)^{i\omega/a}\Gamma
 \left(-\frac{i\omega}{a}\right){\rm e}^{\pi\omega/2a}\left( 1 + \frac{2\theta^{01}\omega}{az^2}\left[\frac{i\omega}{a}
-1\right] \right) \;,\label{calka}
\end{equation}
Hence, it is easy to  deduce the thermal power    spectrum at
negative frequency, which  looks as follows
\begin{eqnarray}
\omega P^{{\rm Twisted}}_{\theta}(-\omega) =\frac{1}{T}\frac{1}{{\rm
e}^{\omega/T}-1}\left( 1 - \frac{2\theta_{01} \omega}{\pi T z^2}
\right)  + {\cal O}(\theta_{01}^2) \;,\label{calka1}
\end{eqnarray}
with Hawking temperature $T=a/2\pi$ associated with the acceleration
of twisted observer.

\section{{{Final remarks}}}
In this article we provide three (linearized) Rindler spaces
corresponding to the canonically, Lie-algebraically and
quadratically twist-deformed Minkowski space-times \cite{chi},
\cite{lie2}. Further, we demonstrated that in the case of canonical
deformation, there  appear   corrections to the Hawking
thermal radiation which are linear in parameter $\theta^{\mu\nu}$.

It should be noted, that the above results can be extended in
different ways. First of all, the complete form of  Rindler
space-times can be find with the  use of  complete  twist
differential operators
\begin{equation}
{\cal Z}_{\cdot}^{-1} = \exp {\cal O}_{\cdot}(z,\partial_z)\;,
\label{complete}
\end{equation}
which appear respectively in the  formulas (\ref{rstar}),
(\ref{rstar2}) and (\ref{rstar3}). However, due to the complicated
form of  operators ${\cal O}_{\cdot}(z,\partial_z)$ such a problem
seems to be  quite difficult to solve from technical point of view.
Besides, one can provide the quantum Rindler spaces associated with
the so-called generalized Minkowski space-times
\begin{equation}
[\;{ x}_{\mu},{ x}_{\nu}\;] = i\theta_{\mu\nu} +
i\theta_{\mu\nu}^{\rho}x_{\rho}\;, \label{general}
\end{equation}
investigated recently in the series of papers
\cite{lulya}-\cite{genpogali}. The studies in these   directions are
already  in progress.

\section*{Acknowledgments}
The author would like to thank J. Kowalski-Glikman and J. Lukierski
for valuable discussions.\\
This paper has been financially supported by Polish Ministry of
Science and Higher Education grant NN202318534.

\end{document}